\documentclass[12pt,preprint]{aastex}

\shorttitle{A Study of PSR B1951+32 in CTB 80}
\shortauthors{Moon et al.}

\begin{document}

\title{PSR B1951+32: A Bow Shock-Confined X-ray Nebula, 
       a Synchrotron Knot and an Optical Counterpart Candidate}

\author{D.-S. Moon\altaffilmark{1,2,3},
J.-J. Lee\altaffilmark{4},
S. S. Eikenberry\altaffilmark{3,5},
B.-C. Koo\altaffilmark{4},
S. Chatterjee\altaffilmark{3,6}, \\
D. L. Kaplan\altaffilmark{7},
J. J. Hester\altaffilmark{8},
J. M. Cordes\altaffilmark{3},
Y. A. Gallant\altaffilmark{9,10},
L. Koch-Miramond\altaffilmark{10}}

\altaffiltext{1}{Robert A. Millikan Fellow, Division of Physics, Mathematics and Astronomy,
Caltech, MC 103-33, Pasadena, CA 91125}
\altaffiltext{2}{Space Radiation Laboratory, Caltech, MC 220-47, Pasadena, CA 91125; moon@srl.caltech.edu}
\altaffiltext{3}{Department of Astronomy, Cornell University, Ithaca, NY 14853; cordes@astro.cornell.edu}
\altaffiltext{4}{Astronomy Program, SEES, Seoul National University, Seoul 151-742, Korea; jjlee@astro.snu.ac.kr
, koo@astrohi.snu.ac.kr}
\altaffiltext{5}{Department of Astronomy, University of Florida, Gainesville, FL 32611; eiken@astro.ufl.edu}
\altaffiltext{6}{Jansky Fellow, National Radio Astronomy Observatory, Socorro, NM 87801; schatter@aoc.nrao.edu}
\altaffiltext{7}{Department of Astronomy, Caltech, MC 105-24, Pasadena, CA 91125; dlk@astro.caltech.edu}
\altaffiltext{8}{Department of Physics and Astronomy, Arizona State University, Tempe, AZ 85287; jhester@asu.edu
}
\altaffiltext{9}{Groupe d'Astroparticules, Universit\'e Montpellier II, 34095 Montpellier Cedex 5, France; galla
nt@gamum2.in2p3.fr}
\altaffiltext{10}{DAPNIA/Service d'Astrophysique, CEA/Saclay, 91191 Gif-sur-Yvette Cedex, France; lkoch@discover
y.saclay.cea.fr}

\begin{abstract}
The radio pulsar~B1951+32 and the supernova remnant CTB~80 provide
a rich laboratory for the study of neutron stars and supernova remnants.
Here, we present ground-based optical and near-infrared observations
of them, along with X-ray observations with {\em Chandra}
and a re-analysis of archival data obtained with the {\em Hubble Space Telescope}.
The X-ray observations reveal a cometary pulsar wind nebula which appears to be
confined by a bow shock produced by high-velocity motion of the pulsar,
making PSR~B1951+32 a rare pulsar exhibiting both an H$\alpha$ bow shock and
a shocked X-ray pulsar wind nebula.
The distribution of H$\alpha$ and radio continuum emission is indicative
of a contact discontinuity of the shocked pulsar winds and shocked
ambient medium at $\sim$0.05 pc.
On the other hand, the optical synchrotron knot of PSR~B1951+32
likely has a flat spectrum in the optical and near-infrared wavebands,
and our astrometry is consistent with only one of the two
reported optical counterpart candidates for the pulsar.
\end{abstract}

\keywords{pulsars: individual (PSR~B1951+32) --- shock waves
--- stars: neutron --- supernova remnants: individual (CTB~80)}

\section{Introduction}

Rotation-powered pulsars (RPPs) exhibit diverse interesting phenomena
ranging from periodic pulsations to supersonic motions through
the interstellar medium, as well as relativistic winds often revealed
by interaction with the neighborhood.
More than 30 years after the discovery of RPPs,
however, the origins for many of these phenomena still remain out of our
grasp, although there has been a recent surge of development in some
of these fields, largely driven by new X-ray satellites typified by {\em Chandra}.
One of the many unprecedented features of the new X-ray satellites
is superb angular resolutions; for example, {\em Chandra} gives $\lesssim$ 1$''$
accuracy in the positions of X-ray sources, comparable to the typical
optical and near-infrared (IR) observations.
This enables high-precision comparative studies of X-ray and optical/near-IR
results, such as more robust searches for optical/near-IR counterparts
to RPPs and multi-wavelength studies of their close vicinity.

In many ways, the $\sim$39.5 ms radio pulsar PSR~B1951+32 \citep{kcb+88}
in the CTB~80 supernova remnant (SNR) provides a rare
opportunity to study the interesting phenomena of RPPs.
It is a moderately young pulsar of 1.1 $\times$ 10$^5$ yrs spin-down age,
which is comparable to, but slightly larger than, the age determined by its proper motion
\citep[6.4 $\times$ 10$^4$ yrs;][]{mgb+02} and the dynamical age of  CTB~80
\citep[7.7 $\times$ 10$^4$ yrs;][]{krh+90}, indicating a finite initial spin.
The measured proper motion ($25 \pm 4$~mas~yr$^{-1}$)
corresponds to a velocity of $240 \pm 40$~km~s$^{-1}$ at 2 kpc,
moving away from the SNR center.
In the optical, PSR~B1951+32 is located in a $\sim$$1\arcmin$ nebular
core emitting both Balmer-dominated and forbidden lines.
\citet{hk88, hk89} suggested that the core is a bow shock formed by the relativistic
winds from the pulsar when it encountered the material behind a
radiative shock of the SNR.
{\em ROSAT} X-ray observations were also suggestive of the pulsar
wind nature of the core, albeit its inadequate spatial resolution \citep{sof95}.
On the other hand, using archival data from the {\em Hubble Space Telescope} (HST),
\citet{hester00} reported a possible optical synchrotron knot near PSR~B1951+32, and
\citet{bgs02} claimed the detection of its two optical counterpart candidates.
However, the lack of color information and rather
large astrometric uncertainties make it difficult to reach any
further conclusion on their nature.

In this {\em Letter}, we present X-ray and optical/near-IR observations of
PSR~B1951+32, searching for the predicted X-ray pulsar wind nebula in the
optical core, as well as the emission from the claimed optical synchrotron
knot and counterpart candidates.

\section{Observations and Results}

The CTB 80 core was observed with {\em Chandra} on 2001 July 12
for a total good time of 78.2 ks using its Advanced CCD Imaging Spectrometer (S3 chip),
with a custom sub-array mode of 192 rows, 
minimizing pile-up on the pulsar while still imaging the entire core. 
Figure~1 shows a resulting X-ray image, where a strong point source 
(i.e., PSR~B1951+32; $\sim$21,900 photon counts) is embedded in diffuse emission.  
Its position is (19:52:58.23, +32:52:41.0) with 0\farcs02
uncertainties\footnote{All positional uncertainties quoted in this 
{\em Letter} represent the 90\% confidence levels.} in each coordinate (J2000).
Considering the 0\farcs6 systematic uncertainty of the {\em Chandra} astrometry,
the X-ray position is consistent with that inferred from pulsar radio timing \citep{flsb94}.
Additionally, several other point sources were detected in the {\em Chandra} image,
and Table~1 lists four of them which have isolated counterparts in the 2MASS point source catalog.
The offsets between their {\em Chandra} and 2MASS coordinates are smaller than 0\farcs7, 
comparable to the combined systematic uncertainty of the {\em Chandra} and 2MASS astrometry.
(The systematic uncertainty of the 2MASS astrometry is 0\farcs2.)

We observed PSR~B1951+32 on 2001 July 16 with the  $2048 \times 2048$ pixel CCD
of Carnegie Observatories Spectrograph and Multi-object Imaging Camera (COSMIC)
on the 5-m Palomar Hale telescope.
We used $g$ (500 nm), $r$ (655 nm), and $i$ (800 nm) broad-band 
filters with 100~s integration time for each band, 
and proceeded with the standard data reduction,
such as bias/dark-subtractions and flat-fielding.
We next used $\sim$100 2MASS point sources and the IRAF task {\em ccmap}
to obtain 0\farcs28 (RA) and 0\farcs40 (DEC) astrometric uncertainties of the COSMIC images.
We also performed $K_{\rm s}$ band (2.15~$\mu$m) observations
with the Keck~I telescope on 2003 November 6 using Near-IR Camera (NIRC) 
equipped with a $256 \times 256$ pixel InSb detector.
We obtained 18 dithered frames, each with 42~s integration time,
and subtracted dark and median-combined sky frames.
We then shifted and combined them to make a final image.
In addition, we re-analyzed archival data for CTB~80
from Wide Field and Planetary Camera~2 aboard HST on 1997 October 2.  
We used $\sim$10 2MASS point sources to obtain the astrometric solutions 
for the NIRC and HST images with IRAF {\em ccmap}.

Figure~2 compares the H$\alpha$ (HST F656N band; 2.2~nm wide at 656.4~nm) 
image of the CTB~80 core with the X-ray image.
In order to match them, we undertook the following procedure.
{\it First}, we tied the {\em Chandra} image to the COSMIC images using the point sources in Table~1, 
which gave 0\farcs25 uncertainty in each coordinate
(see the following paragraph for a detailed description).
{\it Next}, we replaced the COSMIC image with the HST F656N band image.
The astrometric discrepancy of them was estimated to be 0\farcs24; 
therefore, the two images in Figure~2 were matched with 0\farcs35 uncertainty.
In the bottom panel, we identify a $\sim$$40\arcsec$
cometary X-ray nebula, elongated along the proper motion of
PSR~B1951+32 with the pulsar at the head.
The tail shows diffuse and extended emission in the northeast, 
where the pulsar has traveled.
The south-western boundary of the nebula appears to be confined by a bow shock-like feature in H$\alpha$,
and the X-ray emission shows a steep gradient in the confined region.
Figure~3 presents the distribution of the X-ray and H$\alpha$ emission
along the proper motion of PSR~B1951+32 (i.e., the arrow in the bottom panel of Figure~2),
showing that the H$\alpha$ emission peaks $\sim$7$''$ 
ahead of the pulsar in the direction of its proper motion.

Figure~4 shows the COSMIC~$gri$, HST~F547 (48.6 nm wide at 547.9 nm), 
and NIRC~$K_{\rm s}$ band images centered on PSR~B1951+32.  
In the COSMIC images, we determined the pulsar position in two different ways.
{\em First}, we simply found its {\em Chandra} position in the COSMIC astrometry
with 0\farcs53 (RA) and 0\farcs60 (DEC) uncertainties, 
dominated by the systematic astrometric uncertainties of {\em Chandra} and 2MASS
and by uncertainties in the COSMIC astrometry.
The large dotted circle in Figure~4(f) represents 
the position of PSR~B1951+32 determined in this way.
{\em Secondly}, the {\em Chandra} image was tied to the COSMIC images 
using the X-ray point sources and their 2MASS counterparts in Table~1,
where the (intrinsic) X-ray positional uncertainties of the point sources
and the uncertainties of the COSMIC astrometry dominate the final uncertainties
of 0\farcs25 in each coordinate.
The small outlined circle in Figure~4(f) represents the position of PSR~B1951+32 determined in this way. 
The positions of the two claimed optical counterpart candidates 
(i.e., 1HST and 4HST; Butler et al. 2002) are shown in Figure~4(f) --
while 1HST is within the error circle, 4HST is outside.

The optical synchrotron knot of PSR~B1951+32 reported by \citet{hester00}
was identified in the HST F547M band image in Figure~4(e).  
Given E(B$-$V) $\simeq$ 0.8 extinction towards PSR~B1951+32 \citep{hk89},
the extinction-corrected magnitude of the knot is $20.1 \pm 0.2$ 
(= $[3.3 \pm 0.7] \times 10^{-5}$~Jy) in the F547M band. 
The knot was also clearly detected in the $r$ and $K_{\rm s}$ bands (Figure~4b,d), 
with estimated extinction-corrected magnitudes of 
$20.0 \pm 0.2$ (= $[4.5 \pm 1.4] \times 10^{-5}$~Jy) and $18.1 \pm 0.2$
(= $[3.9 \pm 0.7] \times 10^{-5}$~Jy), respectively.  
(The quoted uncertainties represent the 68.3\% confidence levels.) 
However, emission associated with the knot was not detected in the HST
data obtained with narrow-band line filters of 
F502N (2.7 nm wide at 502.3 nm), F656N, and F673N (4.7 nm wide at 673.2 nm).

\section{Discussion and Conclusions} 

Previous optical and X-ray observations showed that PSR~B1951+32 has likely been interacting with 
the recombining material behind the radiative shock of the CTB~80 SNR, 
likely forming a bow shock in the core \citep[e.g.,][]{hk89,sof95}.
The radio continuum images resemble our {\it Chandra} image,
together with a feature reminiscent of a bow shock \citep[e.g.,][]{mgb+02}.
In Figure~2--3, the H$\alpha$ emission overlaps the X-ray emission in the core 
(especially in the eastern part), 
and, around PSR~B1951+32, it also shows a bow shock-like feature confining the cometary X-ray nebula.
One simple interpretation of the overall emission of the CTB~80 core 
(except for the bow shock-associated features around PSR~B1951+32; see below)
is that the X-ray and radio emission represent the synchrotron radiation 
of relativistic pulsar winds, while the H$\alpha$ emission does 
cooling, recombining thermal plasma shocked by them \citep[e.g.,][]{hk89}.
Note that this still allows the existence of H$\alpha$ emission purely associated with the CTB~80~SNR, 
which might be responsible for the H$\alpha$ emission in the western lobes of the CTB~80 core 
apparently lacking the X-ray and/or radio counterparts, 
different from one in the eastern part.

Obviously, the most distinctive new feature of the CTB~80 core 
is the cometary X-ray nebula headed by PSR~B1951+32 along its proper motion, 
seemingly confined by the H$\alpha$ emission forming 
bow-shock morphology at $\sim$$7\arcsec$ ahead of the pulsar (Figure~2).
The revelation of such a feature is in good accordance with the previous interpretation
of the CTB~80 core \citep[e.g.,][]{hk89},
in which the cometary X-ray nebula represents the shocked pulsar winds 
confined by an H$\alpha$ bow shock formed by collisional excitation 
of the ambient medium via supersonic motion of PSR~B1951+32.
For this, it is important to note that there is a significant contribution
of collisional excitation to the H$\alpha$ emission in the CTB~80 core
\citep[in addition to the recombination mentioned above;][]{hk89},
which is supportive of the bow-shock interpretation.
One interesting result is that the distance (from the pulsar) to the H$\alpha$ bow shock 
($\sim$$7\arcsec$) is larger than the value ($\sim$$3\arcsec$)
obtained for the radio bow shock \citep{mgb+02,cc02},
indicating the location of a contact discontinuity of the inner and outer shocks
between them, with the radio emission representing the inner shock (or shocked pulsar winds).
Assuming $5\arcsec$ to be the projected angular distance to the contact discontinuity,
we obtain $\sim$0.05 pc at 2 kpc.

In order to have the bow shock formed, ram pressure balance is required between the
relativistic pulsar winds and the ambient medium:
$\rho_a v_{\rm psr}^2$ $\simeq$ $\dot E$/$4 \pi \Omega c r_{\rm s}^2$,
where $\rho_a$ is the ambient medium density, $v_{\rm psr}$ the pulsar velocity,
$\dot E$ the pulsar spin-down energy, $4\pi\Omega$ the solid angle through which 
the pulsar winds flow, $c$ the speed of the light, 
and $r_{\rm s}$ the stagnation radius.
Assumption of equipartition between the pressure of magnetic field
and of particle near the bow shock leads to
$B_{\rm eq} \rm (\mu G)$ $\sim$ 50($n_{\rm H}$/cm$^3$)$^{1/2}$($v_{\rm psr}$/100~km~s$^{-1}$),
where $n_{\rm H}$ is the hydrogen number density,
and also to $B_{\rm eq} \rm (\mu G)$ $\sim$ 
200$\Omega^{-1/2}$($\dot E$/10$^{36}$~ergs~s$^{-1}$)$^{1/2}$($r_{\rm s}$/0.01~pc)$^{-1}$.
The observed values $v_{\rm psr}$ $\simeq$ 240~km~s$^{-1}$, 
$\dot E$ $\simeq$ 3.7$\times$10$^{36}$~ergs~s$^{-1}$,
and $r_{\rm s}$ $\simeq$ 0.03~pc ($\simeq$$3\arcsec$) of PSR~B1951+32, as well as $\Omega$ $<$ 1,
estimate that $B_{\rm eq}$ $>$ 100 $\mu$G. This reconciles with \citet{hk89}
which estimated the preshock density and magnetic field to be 
$\sim$50~cm$^{-3}$ and $\sim$600~$\mu \rm G$.
For $B_{\rm eq}$ = 600~$\mu \rm G$, $\Omega$ $<$ 0.1, 
corresponding to highly anisotropic pulsar winds.
If we assume that $B$ $>$ 100~$\mu \rm G$ in the entire cometary X-ray nebula,
its synchrotron cooling time is 
$t_{\rm sync} \rm (yrs)$ $\sim$ 40$E_{\rm keV}^{-1/2}$ ($B/100 \; \mu \rm G$)$^{-3/2}$ $<$ 40 yrs,
where $E_{\rm keV}$ is the observed photon energy in keV.
Considering the size ($\sim$$20\arcsec$) of the cometary X-ray nebula
and the magnitude of the PSR~B1951+32 proper motion ($\sim$25~mas~yr$^{-1}$),
the short cooling time indicates that 
the cometary X-ray nebula has likely been replenished by the relativistic
pulsar winds flowing from the pulsar.
Note that similar results have recently been reported for other pulsar wind nebulae \citep{ket01, get03}.
In conclusion, after the millisecond (recycled) pulsar PSR~B1957+20 \citep{sgk+03},
PSR~B1951+32 is only a second pulsar showing both inner and outer shocks in a pulsar wind nebula,
and it is a unique pulsar exhibiting such
a feature in the optical, X-ray, and radio emission together.

Our results confirm that the claimed optical synchrotron
knot of PSR~B1951+32 is indeed a continuum source in nature.
In Figure~4, the emission from the knot appears to be present in the
all five broad bands (although its significance is weak in the $g$ and $i$ bands).
For the HST data,
the knot is visible only in the relatively line-free F547M band image (Figure~4[e]),
while it is absent in the narrow-band line filter images (i.e., F502N, F656N, and F673N).
Thus, the knot exhibits continuum emission with a flat spectrum
in the optical and near-IR wavebands (see \S~2 for the flux estimation), 
unless it has a significant variability.
The Crab pulsar is also known to have a similar optical synchrotron knot \citep{het+95} 
which might be caused by quasi-stationary shocks from pulsar polar outflows \citep{lou98}.  
On the other hand, of the two optical counterpart candidates for PSR~B1951+32 \citep{bgs02},
our improved astrometry is consistent only with 1HST (while it excludes 4HST; Figure~4).  
However, it is possible that 1HST simply represents non-uniformity 
in the optical synchrotron knot or emission from any background star.
We need multi-epoch, multi-color observations 
to study the synchrotron knot and optical counterpart more thoroughly. 

\acknowledgments
Part of data presented herein were obtained at the W.~M.~Keck Observatory, 
which is operated by the California Institute of Technology (CIT), 
the University of California, and NASA.
This research is based on the data from the archive at STScI,
which is operated by the association of Universities for Research in Astronomy, Inc. 
under the NASA contract NAS 5-26555. 
This research has also made use of the data products from
the Two Micron All Sky Survey, which is a joint project of the
University of Massachusetts and IPAC/CIT, funded by NASA and NSF.
DSM was supported in part by an NSF grant AST-9986898 and 
also by a Millikan fellowship at CIT;
JJL was supported by a KOSEF grant ABRL 3345-20031017.
SSE is supported by an NSF CAREER award AST-0328522;
SC is supported by a Jansky fellowship from NRAO.
DLK is supported by a fellowship from the Fannie and John Hertz Foundation;
YAG was supported in part by EC fellowship HPMFCT-2000-00671.

\clearpage
\begin{deluxetable}{ccccccc}
\tablecolumns{7}
\tablewidth{0pt}
\tablecaption{Parameters of the X-ray Point Sources \label{tbl-1}}
\tablehead{
\colhead{List} & \multicolumn{2}{c}{Coordinates (J2000)\tablenotemark{a}}   & \colhead{{\em Chandra}} & \multicolumn{3}{c}{2MASS Magnitudes}     \\
\colhead{}     & \colhead{{\em Chandra\tablenotemark{b}}} & \colhead{2MASS} & \colhead{Counts}        & \colhead{$J$} & \colhead{$H$} & \colhead{$K_{\rm s}$}   }
\startdata
1   &  (53:11.88[2], 53:11.7[4]) & (53:11.89, 53:11.4) & 17.3 & 12.3    & 12.2    & 12.1 \\ 
2   &  (53:08.21[3], 53:10.3[2]) & (53:08.23, 53:10.3) & 47.0 & 13.2    & 12.7    & 12.6 \\ 
3   &  (52:49.21[2], 52:42.1[3]) & (52:49.22, 52:42.0) & 19.1 & 12.1    & 11.9    & 11.8 \\ 
4   &  (52:47.97[2], 52:47.9[4]) & (52:47.97, 52:47.2) & 24.8 & 13.2    & 12.9    & 13.0 \\ 
\enddata
\tablenotetext{a}{The values are offsets from (19:00:00, +32:00:00).}
\tablenotetext{b}{The numbers in brackets represent statistical uncertainties 
at the 90~\% confidence level in the last quoted digit.}
\end{deluxetable}

\clearpage
\begin{figure}[htf]
\plotone{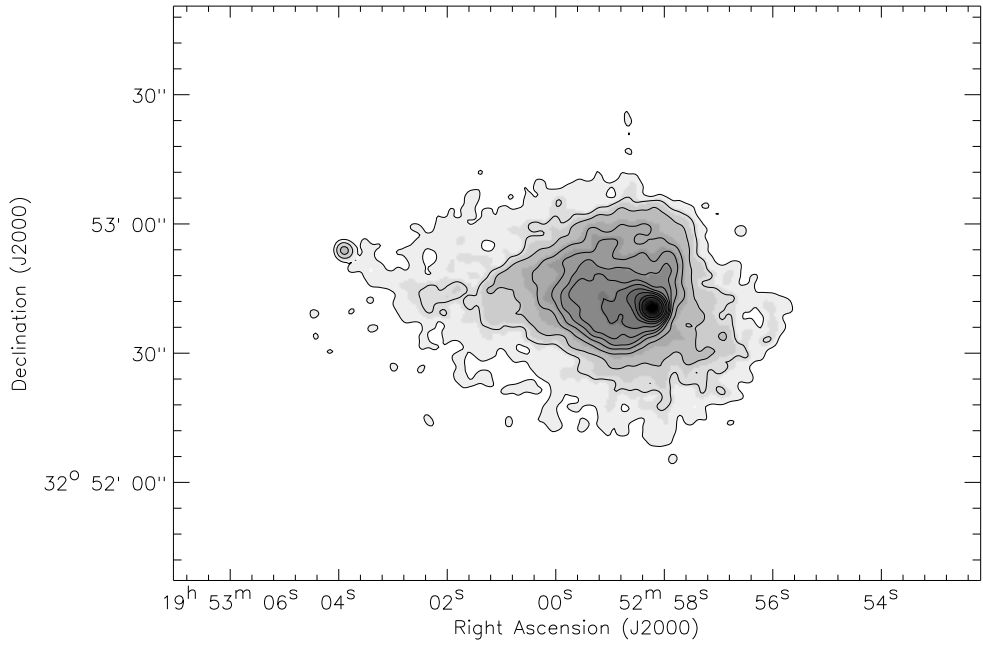}
\caption{{\em Chandra} X-ray (0.3--10 keV) image (both in grey scale and contour)
of the CTB~80 core. The image was convolved with a Gaussian filter of 1$''$ FWHM.
The contour levels correspond to 0.1, 0.2, 0.4, 0.8, 1.1, 1.5, 1.9, 2.6, 3.8,
5.7, 10.5, and 19 \% of the peak brightness.
\label{fig1}}
\end{figure}

\clearpage
\begin{figure}[htbf]
\plotone{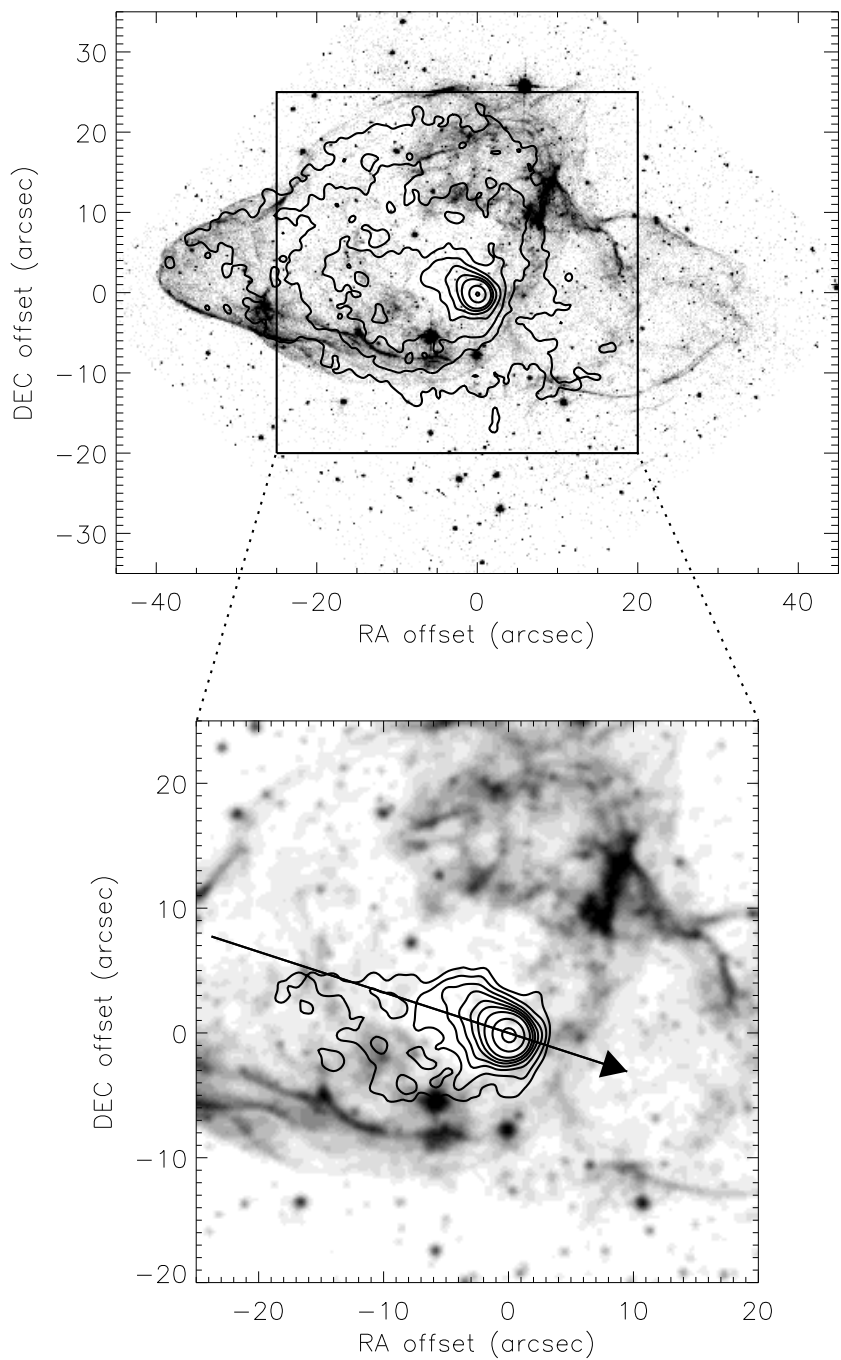}
\caption{{\em Chandra} X-ray image (contour) superimposed on the 
H$\alpha$ image obtained with the HST F656N filter (grey scale). 
The axes represent the offsets from the pulsar position.
The contour levels are 0.1, 0.4, 0.7, 1.3, 2.2, 4.0, 7.3, 36.6, and 95.2 \% of the
peak brightness for the upper panel;
0.9, 1.2, 1.6, 2.2, 3.1, 4.4, 7.3, 22.0, and 73.2 \% for the bottom panel.
The HST image was convolved with a Gaussian filter of 0$''$.4 FWHM.
The arrow in the bottom panel shows the direction of the the 
proper motion of PSR~B1951+32 \citep{mgb+02}.  
\label{fig2}}
\end{figure}

\clearpage
\begin{figure}
\plotone{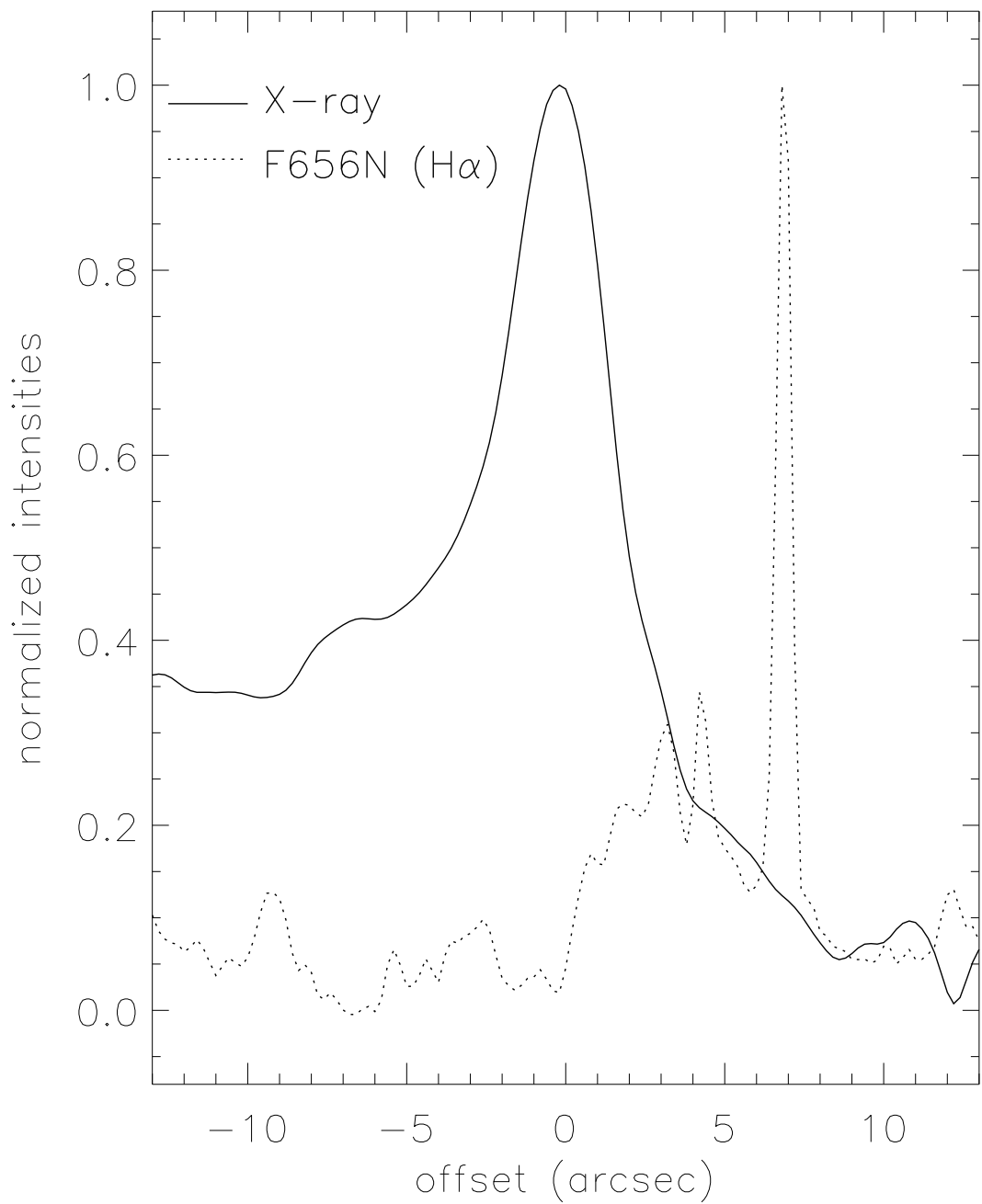}
\caption{Distribution of the normalized intensities of the
X-ray (solid) and H$\alpha$ (dotted) emission in Figure~2
along the proper motion (i.e., the arrow) of PSR~B1951+32.
The X-ray emission is normalized in the logarithmic scale;
the H$\alpha$ emission in the linear scale.
The intensities are calculated by bilinear interpolation
at regularly spaced points on the arrow.
The x-axis represents the offsets from the pulsar,
with negative for the northeast direction.
\label{fig3}}
\end{figure}

\clearpage
\begin{figure}
\plotone{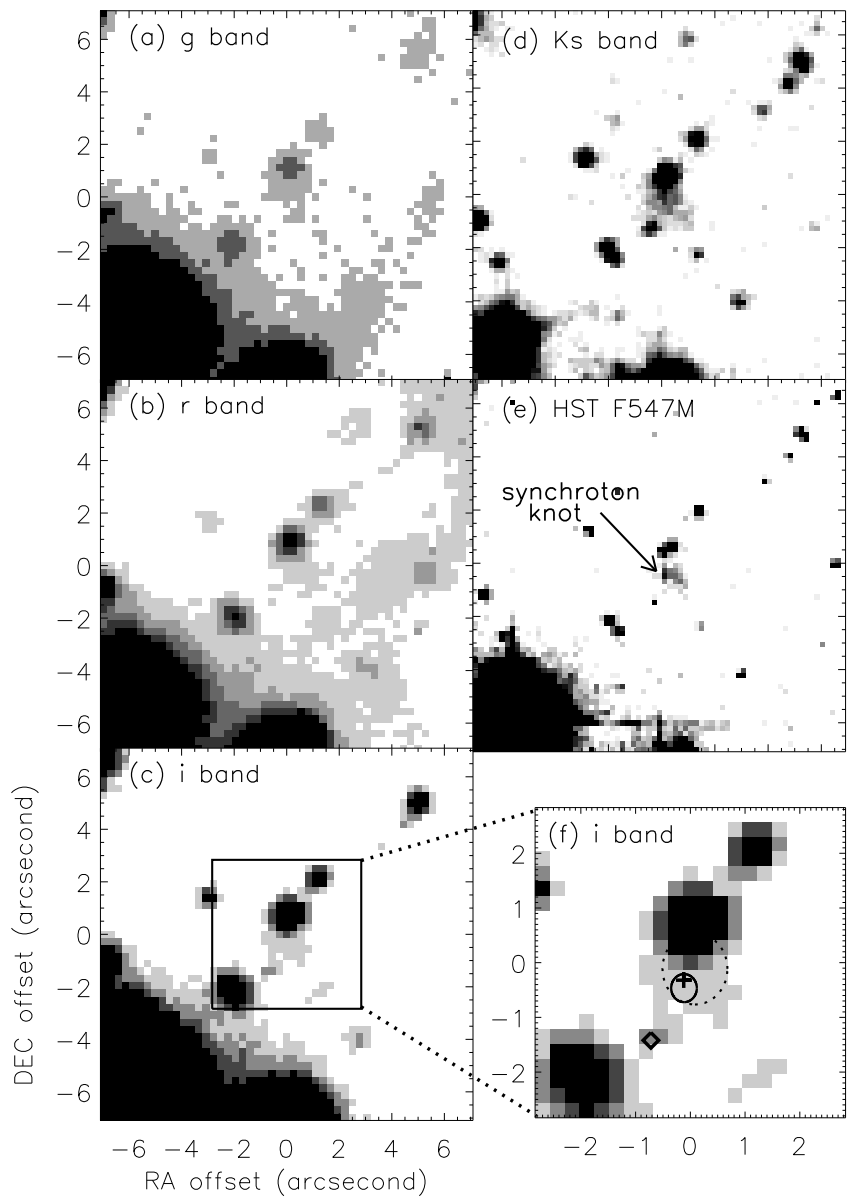}
\caption{(a--c): COSMIC $gri$-band images around PSR~B1951+32.
(d--e): NIRC $K_{\rm s}$ band and HST F547M band images.
The arrow in (e) points to the optical synchrotron knot.
(f): A magnified view of (c). The dotted and outlined circles 
represent the positional uncertainties (90~\% confidence level) of PSR~B1951+32,
determined from absolute astrometry and image alignment, respectively.
The cross and diamond correspond to 1HST and 4HST, respectively.
\label{fig4}}
\end{figure}

\end{document}